\documentclass[dvips,12pt]{article}
\begin{document}
\markboth{}{Motion of a Vector Particle in a Curved Space-time IV}

\title{Motion of a Vector Particle in a Curved Space-time.
IV. Asymptotical shape of caustic}

\author{A.T. Muminov\\
\footnotesize{Ulugh Beg Astronomical Institute,
Astronomicheskaya 33, Tashkent 100052 Uzbekistan}\\
\footnotesize{amuminov2002@yahoo.com}\\
Z.Ya. Turakulov\\
\footnotesize{Ulugh Beg Astronomical Institute,}\\
\footnotesize{Institute of Nuclear Physics, Ulughbek, Tashkent
102132 Uzbekistan}}

\maketitle

\abstract{The studies of influence of spin on a photon motion in a
Schwartzschild spacetime is continued. In the previous paper
\cite{za2} the first order correction to the geodesic motion is
reduced to a non-uniform linear ordinary differential equation and
the equation obtained has been solved by the standard method of
integration of the Green function. If each photon draws a world
line specified by this solution then light rays from infinitely
distant source form a caustic which does not appear without the
spin-gravity interaction. The goal of the present work is to
obtain explicit form of caustic.

}{}{}
\def \vc {\vec}
\newcommand{\T}{\tilde }
\newcommand{\lr}{\left(}
\newcommand{\rr}{\right)}
\newcommand{\lfc}{\left\{}
\newcommand{\rtc}{\right\}}
\newcommand{\lta}{\left\langle}
\newcommand{\rta}{\right\rangle}
\newcommand{\too}{\longrightarrow}
\newcommand{\eps}{\varepsilon}
\newcommand{\vphi}{\varphi}
\newcommand{\om}{\omega}
\newcommand{\fracm}[2]
{{\displaystyle#1\over\displaystyle#2\vphantom{{#2}^1}}}
\newcommand{\sprd}[2]
{\left\langle#1\,,#2\right\rangle\,}
\newcommand{\prt}{\partial}
\def \elmg {electromagnetic }
\def \va {\vc A }
\def \vac {{\vc A }^*}
\def \ca {\bar{A}}
\def \vcx {\dot{\vc x }}
\def \vdx {\delta\vc x }
\def \vca {\dot{\vc A }}
\def \vcca {\dot{{\va\vphantom{A}\,}^*}}
\def \wd {\wedge}
\newcommand{\Ds}[1]{{D#1\over ds}}
\def \dds {\frac{d}{ds}}
\newcommand{\lcom}[1]{{\left[#1\right.}}
\newcommand{\rcom}[1]{{\left.#1\right]}}
\newcommand{\lacm}[1]{{\left\{#1\right.}}
\newcommand{\racm}[1]{{\left.#1\right\}}}
\newcommand{\cc}[1]{\bar{#1}}
\newcommand{\half}{\frac{1}{2}}
\newcommand{\quart}{\frac{1}{4}}
\def \oh {{1/2}}
\def \moh {{-1/2}}
\def \goo {{1-r_g/r}}
\def \gob {(1-r_g/r)}
\def \rd {{r^2-D^2\gob}}
\def \rdi {{r_1^2-D^2(1-r_g/r_1)}}
\newcommand{\ctg}[1]{\tan^{-1}{#1}}
\def \vce {{\vc e }}
\def \vcn {{\vc n }}
\def \cdt {\hspace{0.03em}\cdot\hspace{0.03em}}
\def \fdt {\hspace{0.36em}}
\def \dtx {\dot{x}}
\def \dta {\dot{A}}
\def \dx {{\delta x}}
\newcommand{\gamm}[3]{\,\gamma_{#1#2\cdt }^{\fdt\fdt#3}\,}
\newcommand{\con}[2]{\omega_{#1\cdt}^{\fdt#2}}
\newcommand{\cur}[2]{\Omega_{#1\cdt}^{\fdt#2}}
\def \ccn {\mbox{C.C. }}
\newcommand{\dd}[2]{\fracm{\prt#1}{\prt#2}}
\newcommand{\der}[2]{\frac{d#1}{d#2}}
\def \aRL {\left|\dd{R}{L}\right|}
\newcommand{\sg}[1]{{\varepsilon(#1)}}
\newcommand{\stick}[2]{{\left.#1\right|_{#2}}}

\section{Introduction}

In our recent works \cite{zr2,za2} deviation of photon world line
from a null geodesic was obtained in the first order approximation on
spin-gravity interaction in Schwarzschild space-time. If all world
lines form a parallel beam from an infinitely distant source,
spin-gravitational interaction of photon should give the following
effect. Since the world lines obtained do not lie wholly on one
semi-plane of constant azimuthal angle $\varphi$ they do not
cross the axis of symmetry on which the source is, but pass at some
non-zero distance from it. Then some neighbohood of the back semi-axis
is shadowed by the deviation. The shadow has certain boundary formed by
the envelope of the rays, hence, this surface appears as caustic of
this family of curves. The goal of the present work is to find out the
shape of the surface.

To obtain the shape of the surface we need to consider some
details of calculations of the reference geodesic and our
first-order approximation. Therefore it is useful to recall the
techniques and the denotions used. We use the standard denotions:
coordinates for the Schwartzschild space time are
$\{t,r,\theta,\vphi\}$ and null geodesics lying wholly on the
$\theta=pi/2$ equatorial ``plane'' are presented in the parametric
form\begin{equation}
t=-\dd{R}{E},\quad\vphi=\vphi_0+\dd{R}{L},\quad\theta\equiv\pi/2
\label{eq4geo}\end{equation}where\begin{equation}\label{R}
R(r)=-\sg{t}\int_{r_0}^r\fracm{\sqrt{E^2r^2-L^2(\goo)}}{r(\goo)}dr,
\end{equation}$\sg{t}=\mbox{signum}\,t$ which follows from the
corresponding solution $\Psi=Et-L\vphi+R(r)$ of the
Hamilton-Jacobi equation $<d\Psi,d\Psi>=0$. The constants are
chosen such that under $t=0$ each geodesic reaches the minimal
value of $r$, $r=r_0$. Then under $t\to-\infty$ the geodesic
approaches the starting point at the source and under
$t\to+\infty$ it runs away to the flat asymptotics. We put
$\stick{\vphi}{t\to-\infty}$ $=0$, so that $\vphi_0=$
$\aRL_{r\to\infty}$. It must be noted that unlike action of a
massive particle which can be used as a parameter on the geodesic,
that of massless one $\Psi$ takes constant values on null
geodesics in question. Therefore, $\Psi$ cannot be used as a
parameter on the geodesics.

The form of 0-geodesics determined by squaring (\ref{eq4geo}) is given
in terms of hyperelliptic functions. In fact, we do not need the exact
expressions because in the commonplace approximation we employ:
$r_g\ll D$ where $D=L/E$ stands for the impact parameter, all
expressions simplfy (see, for example, Ref\cite{LD}). The radial part
(\ref{R}) of the action function $\Psi$ can be represented as$$
R(s)=-\sg{s}\int^r_{r_0}\sqrt{\frac{1}{(1-r_g/r)^2}-
\frac{D^2}{r^2(1-r_g/r)}}\hspace{0.8em}dr.$$Following \cite{LD} we
substitute$$r(r-r_g)={r'}^2,\quad\mbox{so that}\quad r'\approx
r-r_g/2.$$ Then the function $R(s)$ becomes$$R(s)\approx-\sg{s}
\int\limits^{r-r_g/2}_{r_0-r_g/2}\sqrt{1+\frac{2r_g}{r'}-
\frac{D^2}{{r'}^2}}\hspace{0.8em}dr'.$$Moreover, we have$${r'}^2+
2r_gr'-D^2=x^2-B^2,\quad B^2=D^2+r_g^2,\quad x=r'+r_g\approx
r+r_g/2.$$Another substitution in the integrand yields:$$
R(s)\approx-\sg{s}E\int^{r+r_g}_{D}\frac{(x^2-D^2)^{1/2}}{x-r_g}\,dx.
$$This result allows to represent the shape of light ray approximately
in analytical form considered below\section{Analytical representation
of the shape of the ray}

We see that the factor annullating the integrand at the lower limit
is separated and now we can safely expand another factor putting
$r_g/x\leq r_g/D\ll1$ that gives:\begin{equation}\label{Rapp}
R(s)\approx-\sg{s}E\int^{r+r_g}_{D}\frac{(x^2-D^2)^{1/2}}{x}
\left(1+r_g/x\right)\,dx.\end{equation}
It is seen that the integral admits analytical representation.

The shape of the null geodesic is determined by dependence of angle
$\vphi$ on $s$ given by eq. (\ref{eq4geo}) due to Hamilton-Jacobi
theorem. Substituting eq. (\ref{Rapp}) into the eq. for $\vphi$ we
obtain:$$\vphi\approx \vphi_0+\eps\int^{r+r_g/2}_D
\left\{\frac{D}{x\sqrt{x^2-D^2}}
+\frac{r_gD}{x^2\sqrt{x^2-D^2}}\right\}dx.$$
The constant parameter $\vphi_0$ can be eliminated by corresponding
choice of initial value of this coordinate: $\vphi(-\infty)=0$
accepted in our previous work \cite{za2}. This choice yields the
following explicit form of $\vphi$ as function of the variable $s$:$$
\vphi(s)\approx \pi/2+r_g/D+\sg{s}\left\{ \arccos{D\over
x}+\frac{r_g\sqrt{x^2-D^2}}{Dx}\right\}_{x=r+r_g/2}.$$
Asymptotical shape of the caustic to be found is formed by $s>0$ parts
of the geodesics:\begin{equation}\label{asphi}\vphi(s>0)\approx\pi/2+
r_g/D+\arccos{D(r-r_g/2)\over r^2}+\frac{r_g\sqrt{r^2-D^2}}{Dr}.
\end{equation}

It was shown in our previous work \cite{za2} that in the first order
approximation the geodesics deviate only in $\prt_\theta$ direction.
The deviation is specified by the vector of the deviation
$\delta\vc x $ $=\delta x^2r^{-1}\prt_\theta$. Its only non-zero
component for the outgoing branch of the trajectory $(s>0)$ has the
following asymptotical behavior under $r\to\infty$:\begin{equation}
\label{dx2}\delta x^2\approx{r_gr\over ED^2}\end{equation} where
we have introduced local Cartesian coordinate $x^2$ such that $dx^2=
rd\theta$. Finally, approximate world line of photon is given by the
reference geodesic and the small vector of deviation just provided.
\section{The shape of asymptotical part of the caustic}

Since the shape of caustic of stationary beam of light is stationary
as well as the Schwarschild space-time itself, we do not need the time
coordinate on the world line of given photon and can consider only the
shape of separate light ray in the space endowed with standard
spherical coordinates $\{r,\vphi,\theta\}$.

The envelope of a beam of such rays is just the caustic to be
found. Besides, axial symmetry of the envelope allows to represent
it by the shape of curve along which the surface crosses
equatorial semi-plane $0\leq\varphi\leq\pi/2,$ $\theta=\pi/2.$ In
other words, to obtain the desired surface it suffices to find out
the distance between envelope and the axis
$\rho^2=r^2\sin^2(\vphi)+ (\delta x^2)^2$ as function of the
coordinate $r$. This task reduces the problem of constructing the
envelope of family of curves on the semi-plane
\begin{equation}\label{2curves}\rho(r,D)=
\sqrt{r^2\sin^2(\vphi)+(\delta x^2)^2},\end{equation}where each
curve is labeled with certain value of the impact parameter $D$.
Value of $\varphi$ is given by (\ref{asphi}).  It is convenient to
use the coordinates $\{r,\rho\}$ in the semi-plane instead of the
standard ones $\{r,\vphi\}$.

Appearance of caustic can be explained geometrically as follows. Due
to the spin-gravitational interaction the rays instead of crossing the
axis pass at distance $\rho$ from it. The rays which are
asymptotically straight lines, under given value of  the impact
parameter $D$ lie on one-sheet hyperboloids whereas the geodesics do
on cones with same axis. Unlike the cones, the hyperboloids of
distinct $D$ constitute a family which has an envelope. In
two-dimensional picture they are hyperbolas whereas the geodesics are
straight lines incident to the axis. The hyperbolas have an envelope
we are constructing. In this section we simplify the task assuming
that the envelope of hyperbolas almost coincides with the curve on
which their points closest to the axis lie. We call this curve
``simplified'' caustic and first, we explore it. Afterwards we show
that in our approximation this curve coincides with the genuine
caustic constructed in classical approach.

Since all the results are obtained under assumption that the value
$r_g/D$ is small, the impact parameter is to be taken sufficiently big
$D\gg r_g$ that allows to obtain only asymptotical behavior of the
caustic. To do it we need explicit form of the expression for the
function $\sin\vphi(r)$ in our approximation. Straightforward
substitution of the equation (\ref{asphi}) yilds:\begin{equation}
\label{sin}\sin(\vphi)\approx D(1/r+r_g/r^2)-2r_g/D \approx {D\over
r}-{2r_g\over D}.\end{equation}It is convenient to represent the
envelope by the curve $r=r_x(D),$ $\rho=\rho(r_x(D),D)$, where
$r_x(D)$ is the value of the coordinate $r$ under which the ray with
impact parameter $D$ crosses the axis. To find it we solve the
equation$$\sin\vphi(r,D)=0$$ which due to equation (\ref{sin}) reduces
to\begin{equation}\label{rx}r_x={D^2\over2r_g}.\end{equation}This
allows to express the variable $\rho$ in the neighborhood of the
crossing point $r=r_x$:\begin{equation}\label{rhod}
\rho=\delta x^2(r_x)={r_g\over ED^2}\cdot{D^2\over2r_g}={1\over2E}=
{\lambda\over4\pi},\end{equation}where $\lambda$ is wavelength. In
other words, asymptotically the envelope becomes a cylinder of
radius which disappears in astrophysical scales.\section{Classical
approach}

Since simplified caustic is too narrow to be observed it is necessary
to obtain the genuine caustic. The shape of the genuine caustic can be
extracted from shape of envelope of curves (\ref{2curves}) on which
the following determinant vanishes:$$0=\left|\begin{array}{cc}
\vspace{0.3em}{\prt\rho\over\prt r}&{\prt\rho\over\prt D}\\
{\prt r\over\prt r}&{\prt r\over\prt D}\\ \end{array}\right|= \left|
\begin{array}{cc}\vspace{0.3em}{\prt\rho\over\prt r}
&{\prt\rho\over\prt D}\\ 1&0\\\end{array}\right|=
-{\prt\rho\over\prt D}.$$ Indeed, if we use the parameter $D$ as a
coordinate instead of $\rho$, each curve is given by $D=const$. Since
the curves have an envelope the Jacobian of the transformation
vanishes on it, consequently, the envelope to be found is zero of the
determinant. Thus, the equation of the envelope can be written as
follows:\begin{equation}\label{eq4env}{\prt\rho^2\over\prt D}=0.
\end{equation}It is seen that under $\delta x^2=0$ no envelope appears
and since $\delta x^2$ $\approx r_gr/ED^2$ the envelope can appear
only at distances of order (\ref{rx}). Referring to the equations
(\ref{dx2},\ref{2curves},\ref{sin}) we have:
$$\rho^2\approx {r_g^2r^2\over E^2D^4}+D^2+{4r_g^2r^2\over
D^2}-4r_gr.$$Now, substituting the expression into the equation
(\ref{eq4env}) we obtain:$$0={\prt\rho\over\prt D}\approx-
4{r_g^2r^2\over D^3}+2D-8{r_g^2r^2\over D^3}.$$ Taking account that
$1/E\sim\lambda\ll D$ we have finally the same result as in the
equation (\ref{rx}):$$r^2\approx{D^4\over4r_g^2},$$for $r$ being the
coordinate of point of intersection of the ray and the axis.
\section{Angular size of the caustic}

Usually, angular size of an object is specified by the angle under
which the rays from it diverge on the axis. In this sense the
caustic has zero angular size because its shape is cylindric.
However, the rays are not parallel on it, consequently, some
caustic must be seen. Below we calculate the angle under which
they pass near the axis. Since the rays are straight near the
crossing point the angle is $\delta\phi=\left.r{\prt\vphi\over
\prt r}\right|_{r=r_x}$. Substituting this into the equation
(\ref{asphi}) to the above expression gives:$$r{\prt\vphi\over
\prt r}\approx {D\over\sqrt{r_x^2-D^2}}\approx{D\over
r_x}={2r_g\over D} =\sqrt{2r_g\over r}.$$The angular size is
doubled angle $\delta\vphi$, namely, $2\sqrt{2r_g/r}$ where $r$ is
coordinate of observer. This result seems to be somewhat
unexpected because angular size of caustic obtained this way does
not depend on the wavelength whereas in the limit of zero
wavelength the caustic does not appear at all.

Since angular size of the caustic obtained this way is valid for all
wavelengths it can be obtained also for geodesics which expose no spin
gravitational interaction. The angle in question apparently if formed
by geodesics focused onto the axis and must be observed because any
telescope detects only direction from which radation comes. Naturally,
radiation in this case comes from directions which consitute a cone
which must be seen as a bright ring. However, this image has nothing
to do with spin-gravitational effect and exposes only the fact of
lensing. As for the spin-gravitational interaction, it does not
produce any special observable effect but, probably, some diffusion of
the bright ring with dependence of the diffusion on the wavelength.
\section{Conclusion}

This is the closing work of the series started with the article
\cite{zr1} in which Papapetrou equation was derived for a massless
spining vector particle. In the work \cite{zr2} derivation was
slightly improved and an attempt was made to obtain an approximate
solution of the equation. The final version of derivation of
Papapetrou equation from Lagrangian of electromagnetic field is
presented in our work \cite{za1}. In our work \cite{za2} it was
shown that the method used in the work \cite{zr2} is valid only on
the first half of the photon world line and another method, valid
on the whole world line is proposed. The new method allows one to
obtain the whole world line as a small deviation from the
reference geodesic.

All this work was completed with the purpose to explain why black
holes appear in observations as bright rings. We hoped to prove that
these rings are images of caustics formed by photons inclined from
geodesics by the spin-gravitational interaction. This would be another
test of general relativity. In fact, it is not so. As was shown above,
caustics produced by this interaction are asymptotically too narrow
and have actually zero angular size, therefore, is not observable.
Instead, there exists ring-shaped image of another nature which has
nothing to do with spin-gravitational interaction. Angular size of
this image does not depend on the wavelength and exposes only focusing
of geodesics. As for the spin-gravitational interaction, it can only
give a subtle spectroscopic effect on the image.
\end{document}